\documentclass[11pt]{article}
\usepackage{moriond,epsfig}

\usepackage{float}

\usepackage{bm}
\usepackage{amsmath}
\usepackage{amssymb}
\usepackage{latexsym}

\def\Journal#1#2#3#4{{#1} {\bf #2}, #3 (#4)}

\def\NPB{{\em Nucl. Phys.} B}

\def\PRL{\em Phys. Rev. Lett.}

\def\SOV{\em Sov. J. Nucl. Phys.}

\def\be{\begin{equation}}
\def\ee{\end{equation}}
\def\bea{\begin{eqnarray}}
\def\eea{\end{eqnarray}}

\begin{document}
\vspace*{4cm}
\title{BOSE-EINSTEIN CORRELATIONS IN CHARGED CURRENT MUON-NEUTRINO INTERACTIONS IN NOMAD}

\author{ RICCARDO ZEI }

\address{Dept. of Physics, University of Siena and INFN,\\
        56 v. Roma, 53100 Siena, Italy}

\maketitle\abstracts{
Bose-Einstein Correlations in one and two dimensions have been studied in
charged current muon-neutrino interaction events collected with NOMAD. In
one dimension the Bose-Einstein effect has been analyzed with the Goldhaber
and the Kopylov parametrizations. The two-dimensional shape of the source
has been investigated in the longitudinal co-moving frame. A significant
difference between the transverse and the longitudinal sizes is observed.}
\section{Introduction}
This short communication is based on my Master's Degree thesis work~\cite{bib:paper}
and on the consequent paper~\cite{bib:paper} done in collaboration with the NOMAD
group in Pisa. Further details can be found in these papers. Here only the main
results and conclusions will be presented. Bose-Einstein correlations (BEC) are
an ``apparent attraction'' in momentum space between identical bosons. This
effect is due to the symmetrization of the quantum mechanical wave function of
two identical bosons with respect to particle exchange. BEC effect is relevant
if and only if these particles are close to each other in momentum space. BEC
in momentum space are related to the spatial dimensions of the production
source. In fact the shape of BEC depends on the spatial and temporal
distributions of the boson source and on its degree of coherence. So studies
of BEC may lead to a better understanding of the dynamics of the particle
interactions yielding like-sign bosons in the final state.
\section{The phenomenology of BEC}
The BEC effect can be parametrized in terms of the two particle correlation
function $R$ defined as:
\begin{equation}
    R(p_1,p_2) = D(p_1,p_2)/D_{0}(p_1,p_2)
\end{equation}
where $p_{1,2}$ are the particle four-momenta, $D(p_{1},p_{2})$ is the measured
two-particle density and $D_{0}(p_{1},p_{2})$ the particle `` reference density''.
BEC are usually parametrized using the Goldhaber parametrization~\cite{bib:Gold}
which assumes a spherical Gaussian density function for the emitting sources:
\begin{equation}
    R(Q) = 1 + \lambda \exp\big(-R^2_G Q^2\big)
\label{eq:gold}
\end{equation}
where $Q^2 = -(p_1 - p_2)^2 = M^2_{\pi\pi}-4m^2_{\pi}$, with $M_{\pi\pi}$ the
invariant mass of the pion pair and $R_G$ the source size. The chaoticity
parameter $\lambda$ measures the degree of coherence in pion production, i.e.
the fraction of pairs of identical particles that undergo interference
($0 \leqslant \lambda \leqslant 1$).

The Kopylov-Pogdoretskii (KP) parametrization~\cite{bib:Kop} corresponds to a
radiating spherical surface of radius $R_{KP}$ with pointlike oscillators of
lifetime $\tau$:
\begin{equation}
    R(Q_t,Q_0) = 1 + \lambda\big[4J^2_1(Q_t R_{KP})/(Q_t R_{KP})^2\big]/\big[1 + (Q_0 \tau)^2\big]
\label{eq:kp}
\end{equation}
where $J_1$ is the first-order Bessel function, $\vec{p} = \vec{p}_1 + \vec{p}_2$,
$\vec{q} = \vec{p}_1 - \vec{p}_2$, $Q_0 = \left| E_1 - E_2 \right|$,
$Q_t = \left| \vec{q} \times \vec{p} \right|/\left| \vec{p} \right|$. This
parametrization is not Lorentz invariant and the variables are calculated in the
c.m. of the hadronic final state.

The shape of the hadronic source can be measured by studying BEC as a function of
the components of the vector $\vec{q}$. This study is performed in the longitudinal
centre of mass system (LCMS) in order to avoid possible effects caused by Lorentz boost.
This reference system is defined for every particle pair as that where
$\vec{p} = \vec{p}_1 + \vec{p}_2$ is perpendicular to the axis defined by the
hadronic jet direction. In the analysis we use the longitudinal
component $Q_{||}$ and the perpendicular one $Q_{\bot}$ to the hadronic jet axis. The
parametrization of the correlation is then performed separately for the two $Q_{||}$
and $Q_{\bot}$ components:
\begin{equation}
    R(Q_{||},Q_{\bot}) = 1 + \lambda\exp\big(-Q^2_{||} R^2_{||} - Q^2_{\bot} R^2_{\bot}\big)
\label{eq:qtql}
\end{equation}
the longitudinal and transverse dimensions of the hadron source being represented
by $R_{||}$ and $R_{\bot}$ respectively.
\section{Analysis and Results}
A critical point of the analysis is the definition of the ``reference density''
$D_0(p_1,p_2)$: it should be identical to $D(p_1,p_2)$ except for the lack of
BEC effect. The common methods used to build the reference sample from data
events are: ``Unlike-sign pairs'', ``Mixed event'' and ``Reshuffled $P_t$''
techniques. We have carefully tested these three methods with a full Monte Carlo
simulation. We have required that, in absence of BEC, the distributions in $R(Q)$
should be flat or have no structure at small $Q$ ($\leqslant 0.2$ GeV) which would
distort the study of $R$. In consequence of this requirement, the ``mixed
event'' and ``reshuffled $P_t$'' methods have been rejected because of the presence
of an excess and a defect at small $Q$ respectively. The unlike-sign sample has
been found to be the most adequate in the BEC region.
BEC effects are then investigated by looking in the data at the following ratio:
\begin{equation}
    R(Q)=\frac{\mbox{``like-sign''pion-pairs}}{\mbox{``unlike-sign''pion-pairs}}=\frac{N_{++}(Q)+N_{--}(Q)}{N_{+-}(Q)}
\end{equation}

The Monte Carlo samples are also used to estimate possible spurious BEC effects from
non-pion contaminations present in the sample. In fact, after the event and
track quality cuts, we have looked for the purity of the track identification.
The tracks investigated are those obtained from the full Monte Carlo simulation.
We have found that the positive and negative samples of particles used for BEC
studies contain respectively $\approx 61\%$ of ${\pi}^+$ and $\approx 77 \%$
of ${\pi}^-$. In the analysis, all secondary charged particles have been assumed
to be pions, unless identified as muons, electrons or protons. We have studied
how the BEC could be changed by these misidentified tracks and by the use of the
unlike-sign sample as a reference. We have found that contaminations in the pions
sample due to kaons and protons produce only a variation of the overall
normalization of the distribution in $R(Q)$ and no distortion of its shape for
$Q \leqslant 0.2$ GeV. Therefore they do not affect the measurement of the
radius of the emitting source. We have observed that the unlike-sign pair
distribution as reference sample has the essential property of reproducing
faithfully the non-BEC distribution of like-sign pairs (the ratio is reasonably flat).
However, it is dangerously affected by meson resonances and by electron-positron pairs
from photon conversions. For this reason the $Q$ intervals $0 \leqslant Q \leqslant 0.04$
GeV ($e^{+}e^{-}$ pairs from photon conversions), $0.3 \leqslant Q \leqslant 0.45$ GeV
($K^0$ decay) and $0.6 \leqslant Q \leqslant 0.825$ GeV ($\rho$ decay) have been
excluded from the analysis.

In order to take into account the contribution due to the long-range correlations
outside the BEC region, we have introduced in Eq.~\ref{eq:gold}, ~\ref{eq:kp}
and ~\ref{eq:qtql} a second degree polynomial factor $(1 + aQ + bQ^2)$. This choice
inevitably affects the results of the BEC analysis and contributes to the systematic
errors on $\lambda$ and the radius parameters. In the literature linear and quadratic
forms have been also used. We have found that the linear parametrization is inadequate
to be used in the present analysis of the experimental data, while the quadratic
parametrization reproduces almost exactly the results of the polynomial form.

The Goldhaber parametrization gives for the radius of the pion emission region the
value $R_G = 1.01\pm 0.05$(\textit{stat})$^{+0.09}_{-0.06}$(\textit{sys}) fm and for the chaoticity
parameter the value $\lambda = 0.40\pm 0.03$(\textit{stat})$^{+0.01}_{-0.06}$(\textit{sys}). Moreover
the BEC parameters are substantially independent of the particle pair charge.

Using the Kopylov-Podgoretskii parametrization yields
$R_{KP} = 2.07\pm 0.04$(\textit{stat})$^{+0.01}_{-0.14}$(\textit{sys}) fm and
$\lambda_{KP} = 0.29\pm 0.06$(\textit{stat})$^{+0.01}_{-0.04}$(\textit{sys}). Also in this case the
results are independent of the particle pair charge.

Performing a two-dimensional analysis in the LCMS frame yields
$R_{||} = 1.32 \pm 0.14$(\textit{stat}) fm and $R_{\bot} = 0.98 \pm 0.10$(\textit{stat}) fm.
We have found an elongation factor $(R_{||} - R_{\bot})/R_{||} \approx 35\%$.
Our measurements confirm the LEP results~\cite{bib:paper} that in the LCMS
reference frame the longitudinal size of the pion source is 30-40\% larger than
the transverse one.
\begin{figure}[hb]
\begin{minipage}{.5\linewidth}
\begin{center}
\centerline{\epsfig{file=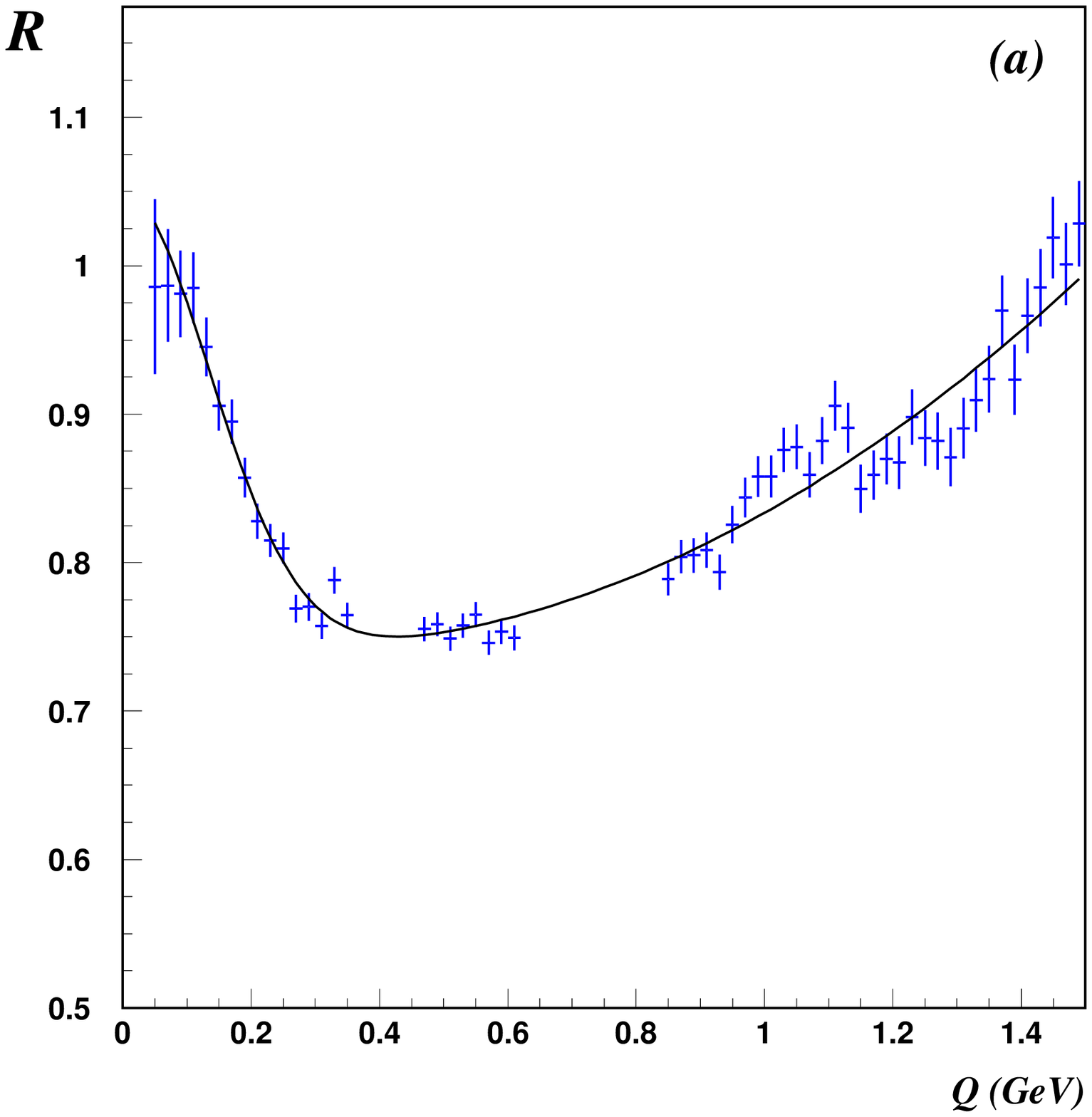,width=0.85\linewidth,angle=0}}
\end{center}
\end{minipage} \hfill
\begin{minipage}{.5\linewidth}
\begin{center}
\centerline{\epsfig{file=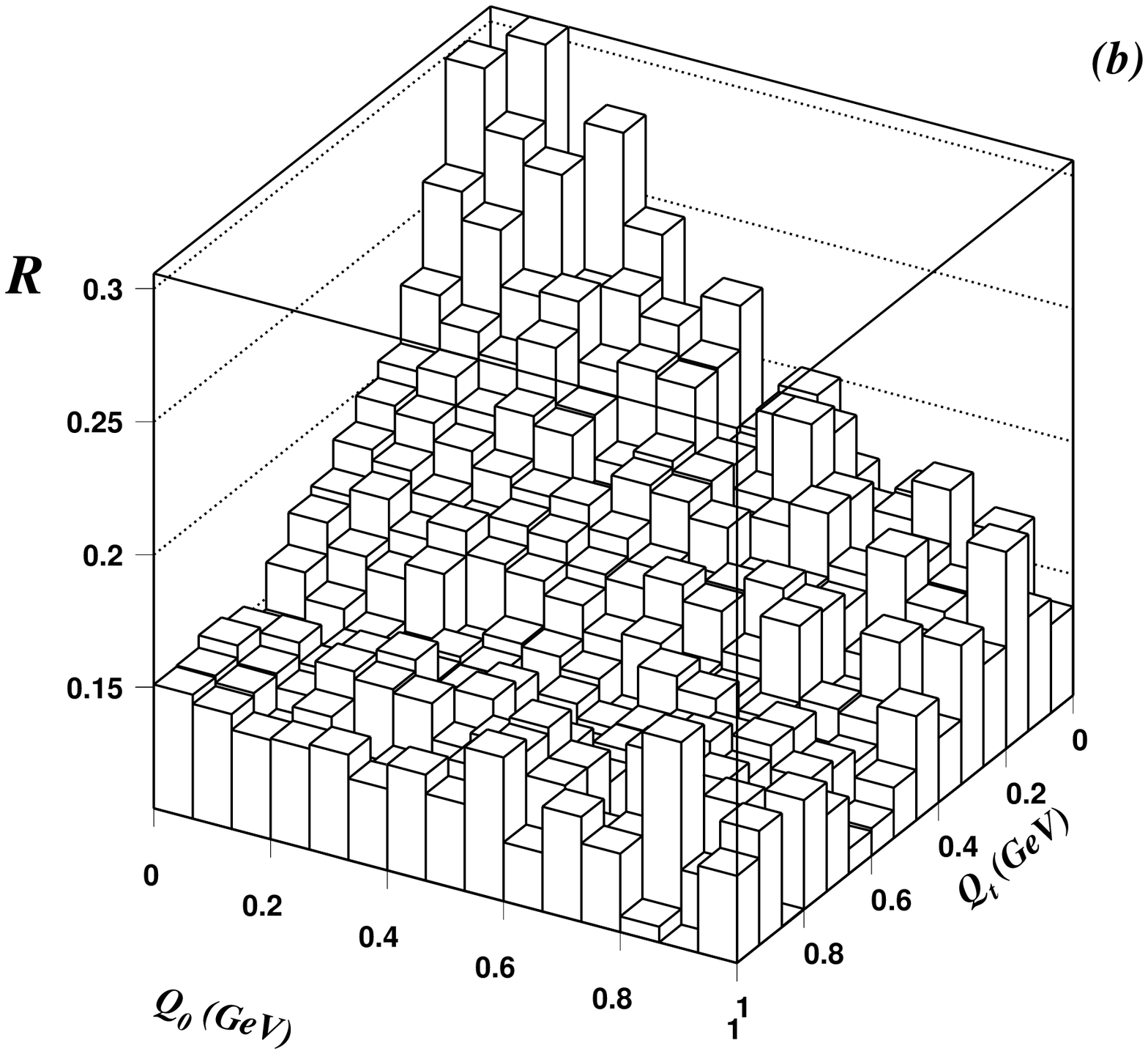,width=0.85\linewidth,angle=0}}
\end{center}
\end{minipage}
\caption{$R$ as a function of the Goldhaber variable $Q$ (a) and of the 
KP variables $Q_t$ and $Q_0$ (b).}
\label{fig:gold_kp}
\end{figure}

Deep inelastic charged current neutrino interactions involve a $d$ quark in the
target nucleon leaving as spectators the remaining quarks. Therefore we could
expect that, at high energy, the two contributions should be fairly well separated
in the c.m. frame of the hadronic jet. But the source radius $R_G$ shows no
differences for particles emitted at different rapidities, demonstrating that the
typical hadronization scale is much longer than the interaction radius, resulting
in a unique hadron source, independent of the detail of the quark interactions.
\begin{figure}[ht]
\begin{minipage}{.5\linewidth}
\begin{center}
\centerline{\epsfig{file=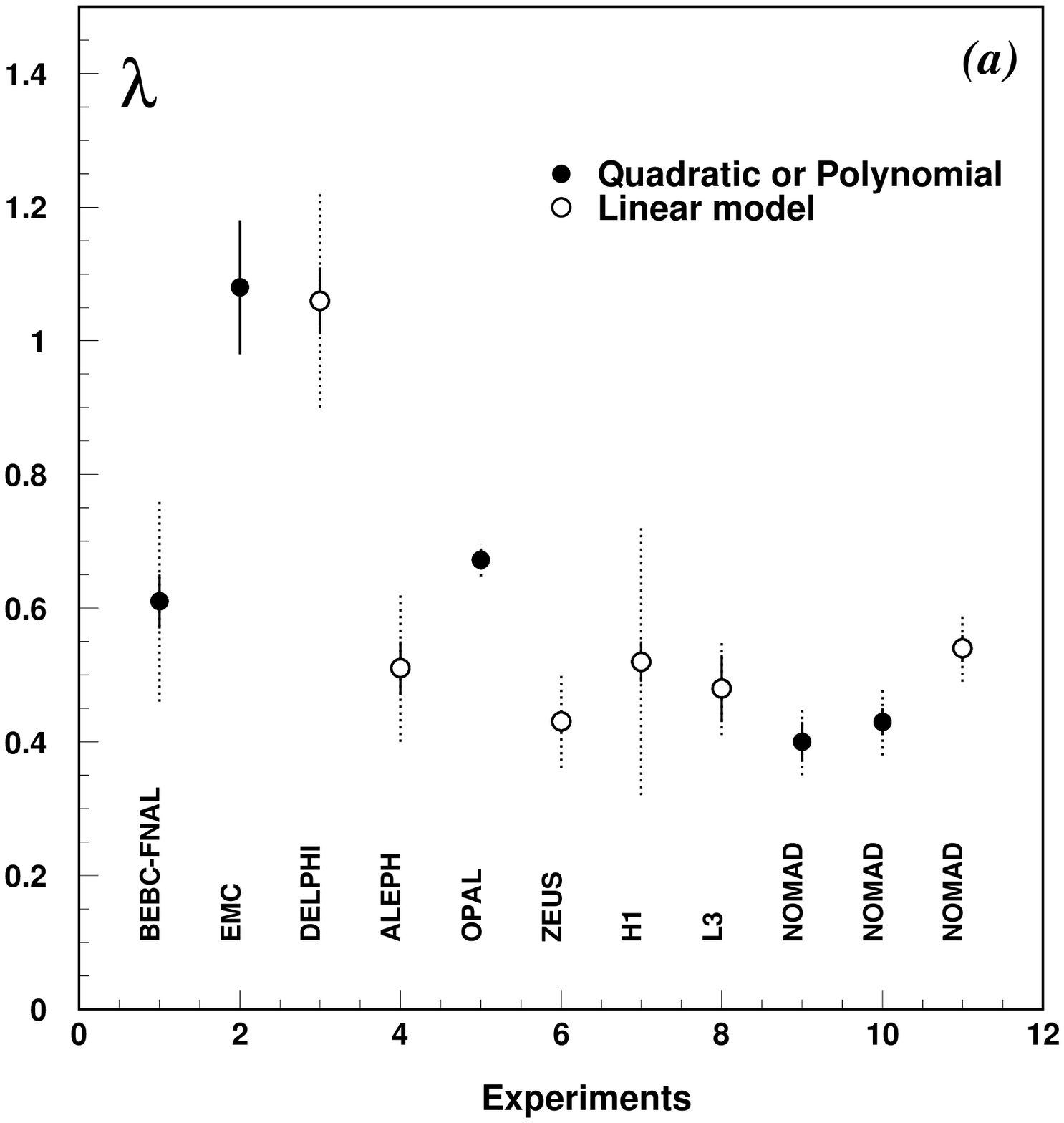,width=0.85\linewidth,angle=0}}
\end{center}
\end{minipage} \hfill
\begin{minipage}{.5\linewidth}
\begin{center}
\centerline{\epsfig{file=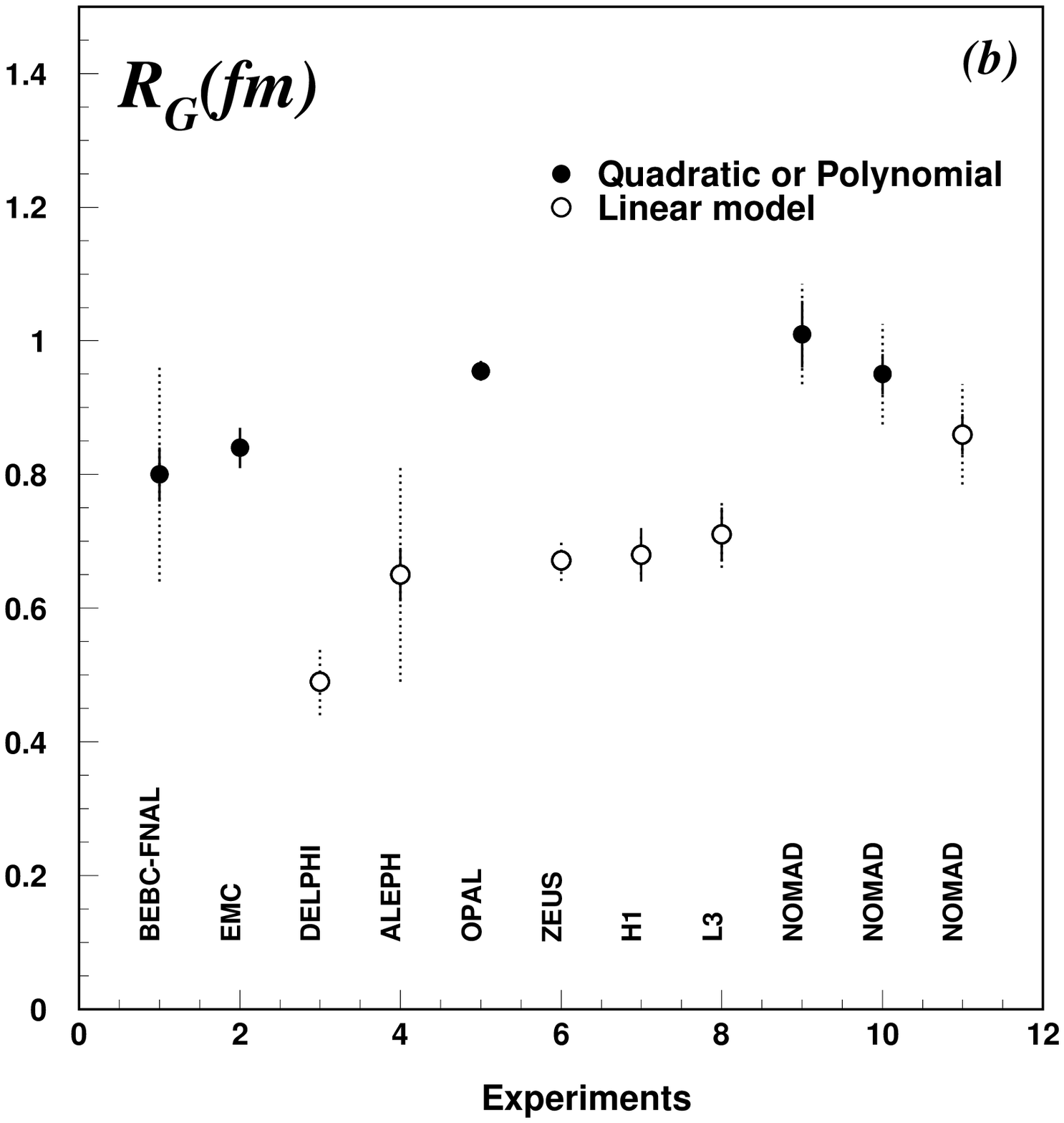,width=0.85\linewidth,angle=0}}
\end{center}
\end{minipage}
\caption{Compilation of results obtained by various experiments for the
chaoticity parameter $\lambda$ (a) and the Goldhaber radius
$R_G$ (b).}
\label{fig:comp}
\end{figure}

In the present analysis, it has not been possible to verify the Goldhaber radius
increase with the event charged multiplicity $N_{ch}$ observed at LEP
experiments~\cite{bib:paper}. This effect is not visible in our data probably
because of NOMAD low multiplicity ($N_{ch} \leqslant 10$) compared to that at LEP
($N_{ch} \gg 10$).

It may be interesting to compare our results with those in the $\pi \pi$ channel
for lepton-induced reactions~\cite{bib:paper}. The results on
$\lambda$ show that there are two groups of experiments ($\lambda \approx 0.5$
and $\lambda \approx 1$) which are consistent within each group, but not between
them. The results on $R_G$ show that the value of $R_{G}$ computed with a linear
model is systematically lower ($R_{G}\approx 0.6$ fm) than the one computed with
a quadratic or polynomial form ($R_{G}\approx 0.9$ fm).
\section{Conclusions}
The size and the chaoticity of the pion source are about 1 fm and about 0.4
respectively, quite independent of the final state rapidity sign of the emitted
pions. A difference of about 35\% is found between the longitudinal and tranverse
dimensions of the source. The final state hadronization processes have universal
features with little dependence on the type or energy of the interacting particles.
\section*{Acknowledgments}
I want to thank Prof. Tarcisio Del Prete, Prof. Vincenzo Cavasinni and the NOMAD 
Collaboration for their strong support and encouragement.

\end{document}